\documentclass[a4paper]{jpconf}
\usepackage{graphicx}
\usepackage{hyperref}
\usepackage{float}

\begin{document}
\title{Deep learning for inferring cause of data anomalies}

\author{V.~Azzolini$^1$, M. Borisyak$^{2,3}$, G.~Cerminara$^4$, D. Derkach$^{2,3}$, G.~Franzoni$^4$, F.~De~Guio$^5$, O. Koval$^{6,3}$, M.~Pierini$^4$, A.~Pol$^7$, F. Ratnikov$^{2,3}$, F.~Siroky$^4$, A.~Ustyuzhanin$^{2,3}$ and J-R.~Vlimant$^8$}

\address{$^1$ Massachusetts Institute of Technology, Cambridge, USA \\
$^2$ National Research University Higher School of Economics,  Moscow, Russia \\
$^3$ Yandex School of Data Analysis, Moscow, Russia \\
$^4$ CERN, European Organization for Nuclear Research, Geneva, Switzerland \\
$^5$ Texas Tech University, Lubbock, USA \\
$^6$ Skolkovo Institute of Science and Technology, Moscow, Russia \\
$^7$ University of Paris-Saclay, Paris, France \\
$^8$ California Institute of Technology, Pasadena, USA}

\ead{fratnikov@hse.ru}

\begin{abstract} 
Daily operation of a large-scale experiment is a resource consuming task, particularly from perspectives of routine data quality monitoring. Typically, data comes from different sub-detectors and the global quality of data depends on the combinatorial performance of each of them. In this paper, the problem of identifying channels in which anomalies occurred is considered.
We introduce a generic deep learning model and prove that, under reasonable assumptions, the model learns to identify 'channels' which are affected by an anomaly. Such model could be used for data quality manager cross-check and assistance and identifying good channels in anomalous data samples.
The main novelty of the method is that the model does not require ground truth labels for each channel, only global flag is used. This effectively distinguishes the model from classical classification methods. Being applied to CMS data collected in the year 2010,  this approach proves its ability to decompose anomaly by separate channels. 
\end{abstract}

\section{Introduction}
Data quality monitoring is a crucial task for every large scale high energy physics experiment.
The challenge is driven by a huge amount of data. A considerable amount of person power is required for monitoring and classification. Previously [1], we designed a system, which automatically classifies marginal cases in general: both of 'good' and 'bad' data, and use human expert decision to classify remaining 'grey area' cases. 

Typically, data comes from different sub-detectors or other subsystems, and the global data quality depends on the combinatorial performance of each of them. In this work, we aim instead to determine which sub-detector is responsible for anomaly in the detector behaviour, knowing only global flag. For this study, we use data [2] acquired by CMS detector [3] at CERN LHC. A proposed system can indicate affected channels and draw the attention of human experts to other channels. Data from the channels, which are reconstructed relying primarily on normally operating sub-detectors, can be used for further specific physics analysis.

\section{Data and feature extraction}
Data preprocessing procedure is the same as in the previous work [1], and described in detailes there. 

All data is divided into time quanta - LumiSections\footnote{Segment of data corresponding to 23 seconds of data taking}, which are labelled as 'good' or 'bad'. Physics object is a proxy for a particle recorded in the detector which is constructed from recorded raw data collected by  several sub-detectors. The information coming from four channels per each used LumiSection. These channels are equivalent to physics objects: muons, photons, particle flow jets (PF) or calorimeter jets (calo). Objects (equivalent to physics particles) are quantiled by their transverse momentum to have fixed number of features in each event. Then every selected object is characterized by its reconstructed physics properties: mass, spatial location, kinematics. And statistics for each feature for the entire lumisection is computed (5 percentiles, mean and variance). 
Additionally, features like instant luminosity, number of particles in event and others were also introduced.

In total we have almost 16k data points (LumiSections) 25\% of which were labeled as anomalous. There are 267 features in muon, 232 in photon, 126 in PF, 266 in calo channel. We use 10\% of data to validate our model during training. And the final test for our system is a computing correlations between the method predictions and experts’ labels for CMS sub-detectors.

\section{Method}
\begin{figure}[h!]
\begin{center}
\includegraphics[width=5in]{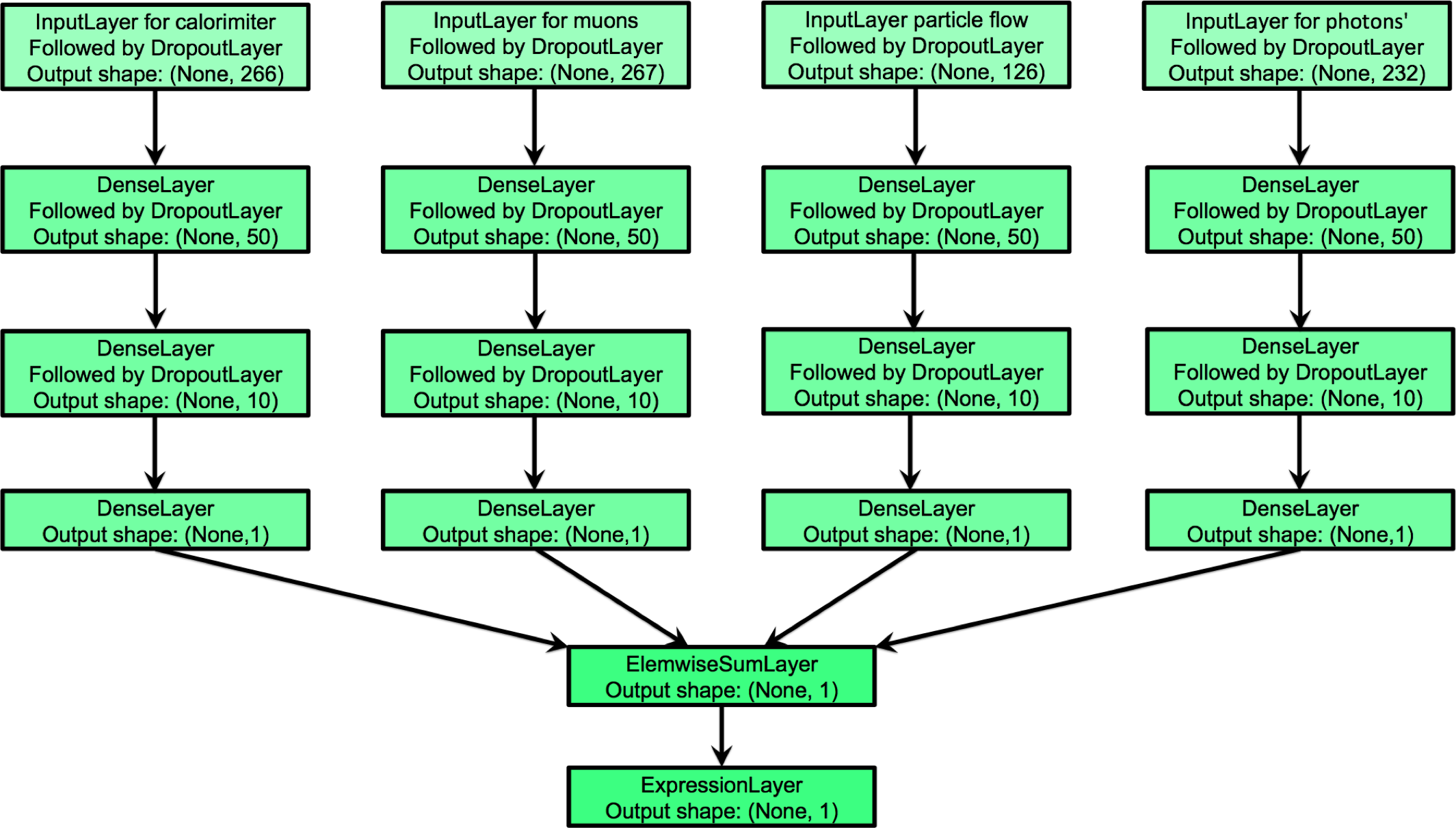}
\end{center}
\caption{\label{fig:nn}NN architecture with four sub-networks for each channel}
\end{figure}

In order to predict a probability of anomaly in different channels separately 
we build a special ‘multi-head’ neural network (NN) configuration presented in Fig. \ref{fig:nn}.

NN consist of four branches and each sub-networks has in input features from 
corresponding channel. Each branch returns a score for its channel. At the end, sub-networks are
connected and the whole network is trained to recover global labels.

In this paper we present results when sub-networks are connected with kind of 'Fuzzy AND' operator: \begin{eqnarray}
\exp\left[\sum_{i=1}^4(f_{i} - 1)\right]
\end{eqnarray}
where $f_{i} - $ is an output of the last layer of sub-networks. 

It is proved for this operator, that under reasonable assumptions, the model learns to identify 'channels' which are affected by an anomaly.\footnote{Proof for 'Fuzzy AND' operator decomposition properties and code of the systems with different operators for network connection are available in \url{https://github.com/yandexdataschool/cms-dqm/} } As network connection operator, logistic regression and min operator with dropout could also be used. 

The main benefit is that proposed approach uses only aggregated global quality tag for training, but allows to predict anomalies for separate channels.

In this way, each subnetwork returns score:
\begin{itemize}
\item close to 1 for good lumisections;
\item close to 1 for anomalies “invisible” from subnetwork’s channel data, but visible to other subnetworks channels in the NN method;
\item close to 0 for anomalies “visible” from subnetwork’s channel data.
\end{itemize}

Thus NN decomposes anomalies by channels.

'Fuzzy AND' approach assumes that there are anomalies not seen from all channels. They look like 'good' data points in the channel, surrounded by them and are classified by this channel as 'good' too. It is desired setting, but it causes a problem of small gradients of 'Fuzzy AND' function for close to the hyperplane data points during training. Long time is needed to change the hyperplane slightly when anomalous data points, which are potentially visible from particular channel, have already 'bad' labels in other channels. A simple cross-entropy loss for 'Fuzzy AND' output of the whole network is not sensitive enough in such cases. To resolve the problem and to accelerate the convergence we use a dynamic loss function:
\begin{eqnarray}
L' = (1-C) \cdot L + C \cdot (L_1 + L_2 + L_3 + L_4)/ 4 ,
\end{eqnarray}
where $L - $cross-entropy loss for 'Fuzzy AND' output of the network; $L_i - $ so called 'auxiliary' losses, cross-entropy of corresponding sub-network scores against global labels; $C - $ decreasing along iterations constant to regulate amount of 'pretraining'.

With such 'soft pretraining' dynamic loss function we can force sub-networks to be more accurate and to take care about ambiguous samples during the first training iterations, but then to pay more attention to the predictive power of the whole NN against global labels. Thus, simple enough separation hyperplane is constructed during training, and problem of the small gradient, which is mentioned above, is avoided.

\section{Results and discussions}
\begin{figure}[H]
\begin{minipage}[h!]{0.47\linewidth}
\center{\includegraphics[width=14pc]{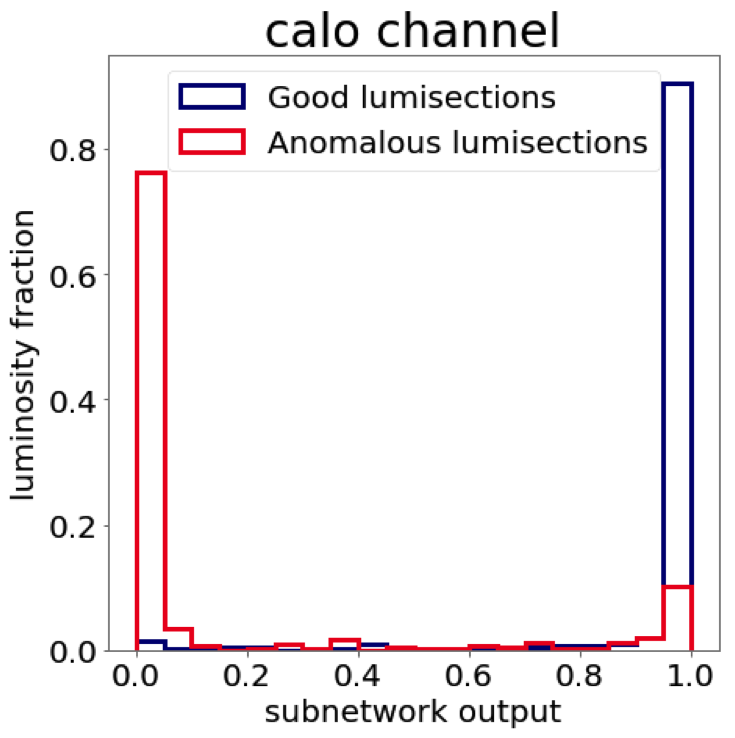}} \\a) 
\end{minipage}
\hfill
\begin{minipage}[h!]{0.47\linewidth}
\center{\includegraphics[width=14pc]{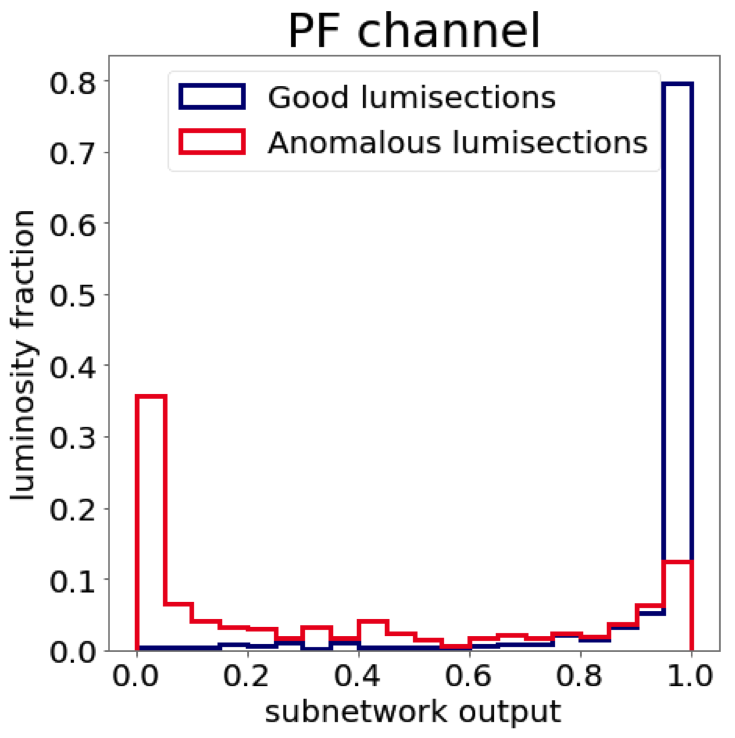}} \\b)
\end{minipage}
\vfill
\begin{minipage}[h!]{0.47\linewidth}
\center{\includegraphics[width=14pc]{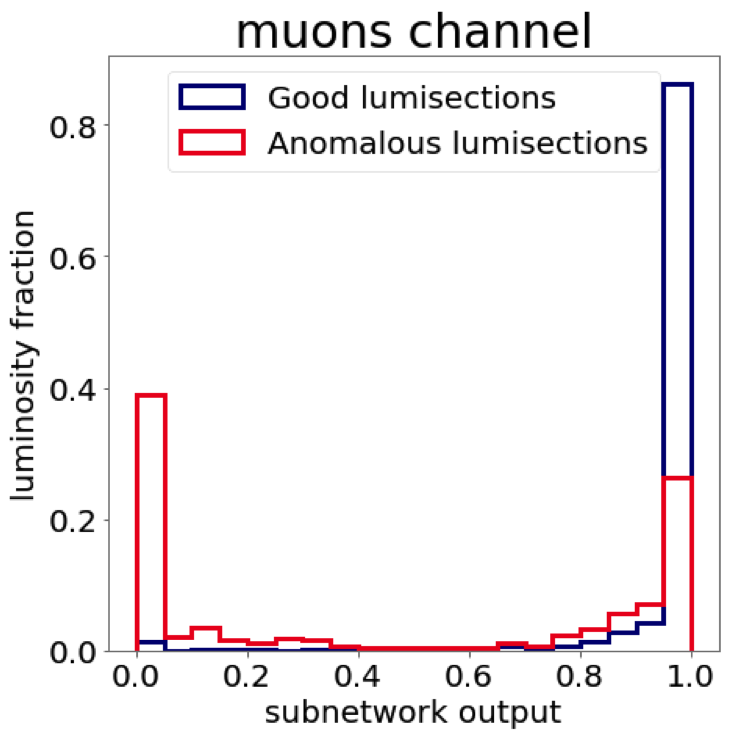}}\\ c) 
\end{minipage}
\hfill
\begin{minipage}[h!]{0.47\linewidth}
\center{\includegraphics[width=14pc]{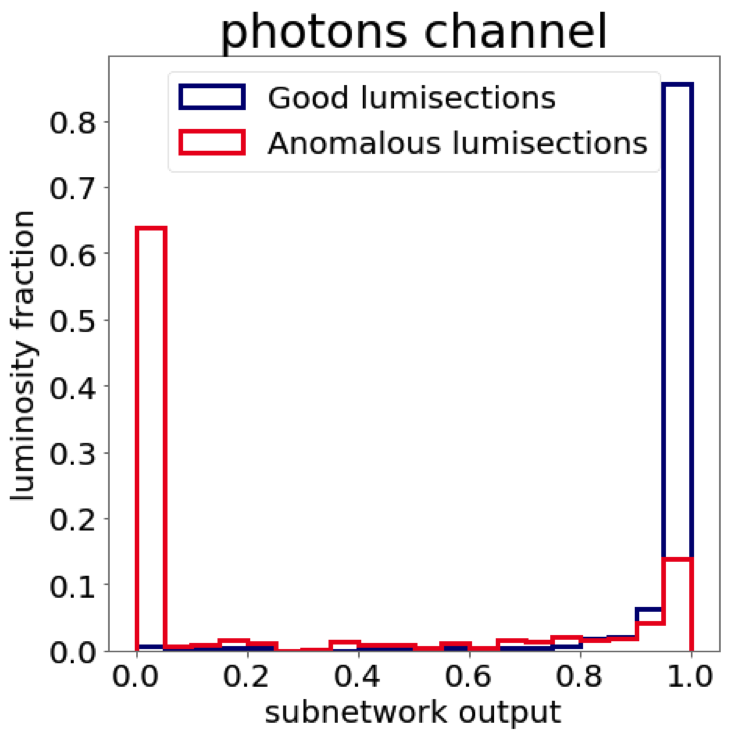}} \\ d) 
\end{minipage}
\caption{\label{fig:predictions} Distributions of predictions returned by NN branches build on features from a) calorimiter, b) particle flow jets, c) muons, d) photons
channels.}
\end{figure}

Being applied to CMS data collected in 2010, method proves its ability to decompose anomaly by separate channels. Distributions of predictions in each NN branch are shown in Fig. \ref{fig:predictions}. As expected, we can see scores close to one for 'good' samples. And 'bad' data has two options, it could be visible from channel (score close to zero) or not. We can think about the second cases, as data is not affected by an anomaly and it is still useful for for further physical analysis.

Thus, method suggests that most of anomalies occurred (or at least better identified) by calo channel. And there are some data from others channels, which does not look like anomalous (predicted labels close to one when the global label is "bad") and can be saved.

\begin{figure}[H]
\begin{center}
\includegraphics[width=5.5in]{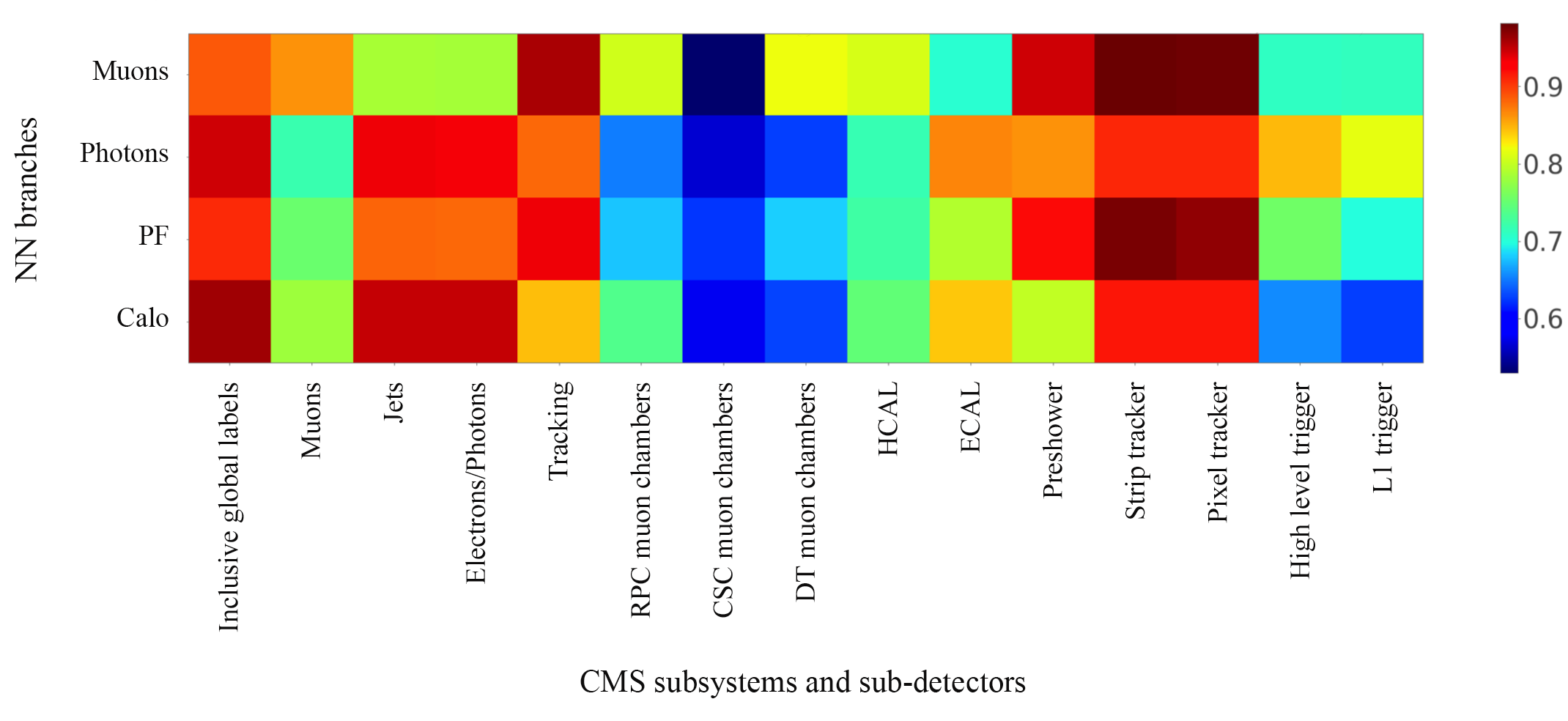}
\end{center}
\caption{\label{fig:matrix}ROC AUC scores of NN branches scores against experts' labels for CMS subsystems}
\end{figure}

In these experiments global predictive power of the whole network on validation set is rather high, ROC AUC score equals to 0.96.
To verify obtained results we calculate correlations between sub-network predictions and experts' labels for CMS subsystems, which were not used for training, as presented in Fig. \ref{fig:matrix}. All ROC AUC scores a higher than 0.5. It means that there is a clear correlation between sub-networks’ outputs and corresponding subsystem labels (for example: Muons vs Tracking) or some of them are almost independent (Photons vs DT muon chambers). But there is no anti-correlations, as it is expected. 

\section{Conclusions}
In this work, we described a deep learning approach for inferring cause of data anomalies. While developed
with the CMS experiment at CERN in mind, we use an agnostic approach which allows the straightforward
adaptation of the proposed algorithm to different experimental setups. Method shows its ability to decompose anomalies by separate channels, being applied to data collected by the CMS experiment in 2010. Although only global quality labels were used for training, we got clear correlation between sub-networks’ outputs and corresponding true subsystem labels, what proves correctness of obtained results.

\section*{Acknowledgements}
The research leading to these results was partly supported by Russian Science Foundation under grant agreement N\textsuperscript{\underline{\scriptsize o}} 17-72-20127.

\end{document}